\documentstyle[12pt]{article}

\begin{document}

{\bf Solar-cycle related variation of solar differential rotation} \\

K. J. Li$^{1,2}$,
X. J. Shi$^{1,3}$, J. L. Xie$^{1,3}$
P. X. Gao$^{1}$
H. F. Liang$^{4}$,
L. S. Zhan$^{5}$, W.  Feng$^{6}$\\
$^{1}$National Astronomical Observatories/Yunnan Observatory,
      CAS, Kunming 650011,   China\\
$^{2}$Key Laboratory of Solar Activity, National Astronomical
Observatories, CAS, Beijing 100012, China\\
$^{3}$Graduate School of CAS, Beijing 100863, China\\
$^{4}$Department of Physics, Yunnan Normal University, Kunming 650093,
 China\\
$^{5}$Jingdezhen Ceramic Institute, Jingdezhen 333001, Jiangxi,
China\\
$^{6}$Research Center of Analysis and Measurement,
Kunming University of Science and Technology, Kunming 650093, China



{\bf Abstract.}
Solar-cycle related variation of differential rotation is investigated through analyzing  the rotation rates of magnetic fields, distributed along latitudes and varying with time at the time interval of August 1976 to April 2008.
More  pronounced differentiation of rotation rates is found to appear at the ascending part of a Schwabe cycle than at the descending part on an average.
The coefficient $B$ in the standard form of differential rotation, which represents the latitudinal gradient
of rotation, may be divided into three parts within a Schwabe cycle. Part one spans from the start to the $4^{th}$ year of a Schwabe cycle, within which the absolute $B$ is approximately a constant or slightly fluctuates.  Part two spans from the $4^{th}$  to the $7^{th}$ year, within which the absolute $B$ decreases. Part three spans from the $7^{th}$ year to the end, within which the absolute $B$ increases.
Strong magnetic fields repress differentiation of rotation rates, so that  rotation rates show less pronounced differentiation, but weak magnetic fields seem to just reflect differentiation of rotation rates. The solar-cycle related variation of solar differential rotation
is inferred to the result of both the latitudinal migration of the surface torsional pattern and the repression of strong magnetic activity to differentiation of rotation rates. \\
{\it Sun: rotation-- Sun:activity-- Sun: surface magnetism}

\section{INTRODUCTION}
The Sun's atmosphere displays differential rotation on its disk: the equatorial region of the Sun rotates
faster than higher latitude regions (26 days at the solar Equator and 30 days at $60^{\circ}$ latitude)
(Howard, Gilman $\&$  Gilman 1984; Sheeley,  Wang $\&$  Nash 1992; Rybak 1994;  Altrock 2003; Song $\&$ Wang 2005;  Le Mouel et al. 2007).
To measure angular rotation velocity of the solar atmosphere, three methods have been mainly used: the tracer method, the spectroscopic
method, and the flux modulation method (Balthasar $\&$ Wohl 1980; Gilman $\&$ Howard 1984;
Brajsa et al.  2000, 2002;   Wohl $\&$ Schmidt 2000; Javaraiah et al.  2005, 2009;
 Wohl et al. 2010; Vats et al 2001; Chandra $\&$ Vats 2011a; Vats, 2012).
Helioseismology measurement can  determine solar rotation
rate in the solar interior (Howe et al. 2000a, 2000b; Antia $\&$  Basu 2001), and the latitudinal shift of
solar rotation rate in solar interior determined by helioseismology measurement (Howe et al. 2009) is  similar to the
torsional oscillation pattern
measured by the spectroscopic method (Howard $\&$ LaBonte 1980; LaBonte $\&$ Howard 1982; Schroter 1987).
Up to now, observations and studies on solar differential rotation have taken  a great achievement (Howard 1984; Schroter 1985; Snodgrass 1992;  Paterno 2010; Vats $\&$ Chandra 2011; Vats 2012).
 However, there are still many aspects, for example, the solar-cycle related and long-term variations of solar rotation rate, unknown at the present (Komm, Howard $\&$ Harvey 1993; Ulrich $\&$ Bertello 1996;  Stix 2002;
 Li et al. 2011a, 2011b).
In this study, we will investigate solar-cycle related variation of solar rotation rate, using data of the rotation rates of magnetic fields, distributed along latitudes and varying with time at the time interval of August 1976 to April 2008, and a new explanation is proposed on such a solar-cycle related variation of the solar rotation rate.

\section{SOLAR-CYCLE RELATED VARIATION OF SOLAR DIFFERENTIAL ROTATION}
\subsection{DATA}
Synoptic magnetic maps indicate the distribution of the magnetic fields on the full
solar surface, and the rotation rates  can be derived.
Chu et al. (2010) employed Carrington-coordinate synoptic magnetic maps, which are produced by the
NSO/Kitt Peak during the year of 1976 to 2003, and by
SOHO/MDI during 2003 to 2008.
They built up a time-longitude stackplot  (McIntosh et al. 1991; Japaridze  et al.  2007) at each
of latitudes $-35^{o}$ to $35^{o}$. On each stackplot there are many tilted magnetic structures, which clearly
reflect the rotation rates, and then they utilized a cross-correlation method to explore the rotation
rates from the tilted structures (Chu et al. 2010). Here, Figure 1 shows the obtained rotation rates distributed along latitudes and varying with time at the time interval of August 1976 to April 2008, namely from Carrington Rotation (CR) 1645 to 2069. In the figure latitudes are negative in  the southern hemisphere, and CRs are translated into calendar years. Velocities  decrease from the solar Equator to high latitudes, and they vary with time, more clearly at relative high latitudes and at the descending part of a sunspot cycle.

Figure 2 show four isopleth lines of  rotation rates, and the corresponding rotation rates are 14.29, 14.25, 13.10, and 13.85 $(^{o}\ day^{-1})$ in turn from low to high latitudes.
Any velocity in the figure is the mean of those at the corresponding latitudes of the two hemispheres.
Shown also in the figure are the minimum and maximum times of sunspot cycles. As the figure shows, rotation rates on the average seem higher when the magnetic field is weak, indicating strong magnetic fields should repress solar differential rotation.

\subsection{SOLAR-CYCLE RELATED VARIATION OF SOLAR DIFFERENTIAL ROTATION}
The solar differential rotation is usually expressed by the
standard formula (Newton $\&$ Null 1951):
$$
\omega (\phi)=A+B sin^{2}\phi
$$
where $\omega (\phi)$ is the solar
sidereal angular velocity at latitude $\phi$, and the coefficients $A$ and
$B$ represent the equatorial rotation rate and the latitudinal gradient
of the rotation, respectively (Howard 1984). At the present, the solar differential rotation for both tracer and spectroscopy
observations is generally expressed by a three-term formula
(Chandra et al. 2010; Song et al. 2011; Takeda $\&$ Ueno 2011; Chandra et al. 2012): $\omega (\phi)=A+B sin^{2}\phi+C sin^{4}\phi$, instead of the traditional two-term formula. However, when the two-term formula is used,  the physical meanings of the coefficients used in the formula can be clearly known: one coefficient represents the equatorial rotation rate, the other,
the latitudinal gradient of rotation, reflecting the differentiation degree of rotation.  Thus it is convenient to
discuss the relationship of differentiation of rotation and solar activity, as does in this study. This is the reason why the two-term is used here. In addition,
Wohl et al. (2010) showed in their Figure 3 that the latitudinal
distributions given respectively by the two- and three-
term formulae are very close to each other at the low
and middle latitudes (less than 45 degrees). Here, rotation rates  at the equatorial range between
latitudes $\pm 35^{0}$ are considered, it is thus sufficient to use just the two-term, and so do Beck (2000),
Chandra et al. (2009), Badalyan (2010), and Chandra $\&$ Vats (2011b).

The latitudinal distribution of rotation rates is fitted by the two-term formula
at each of all considered  CRs.
Figure 3 shows the correlation coefficient of the formula fitting to the distribution of differential rotation rates along latitudes, and it is larger than 0.97 at each CR.  The correlation  is thus
statistical significant at the $99.99\%$ confidential level. The formula can give a very good fitting at each of all considered CRs.
Figure 3 also shows the obtained coefficients $A$ and $B$.  $A$ peaks around the minimum and maximum times of sunspot cycles, implying the existence of a period of about a half Schwabe cycle in length. The absolute value of $B$ on the average seems larger when the magnetic field is weak, indicating the magnetic field should repress the differentiation of solar rotation. A special feature of $B$ is that its absolute value clearly decreases and then increases within the descending part of a sunspot cycle.

Values of the coefficient $B$  are averaged for all considered time within the
same solar cycle phase relative to the nearest preceding sunspot minimum. Resultantly, Figure 4
shows the dependence of the coefficient values on the phase of the solar
cycle relative to the nearest preceding sunspot minimum, and their corresponding standard errors. Different solar cycles have different period lengths,
thus, also shown in the figure are the dependence of the coefficient values  on the phase of the solar
cycle relative to sunspot maximum, and their corresponding standard errors. As the figure displays,
the special feature of $B$ can be clearly seen.

We calculate the average of rotation rates over latitudes of $-35^{o}$ to $35^{o}$, and then
the obtained mean values of rotation rates are averaged for all considered time within the
same solar cycle phase relative to the nearest preceding sunspot minimum. Resultantly, Figure 4
shows the dependence of  the latitudinally mean values of rotation rates on the phase of the solar
cycle relative to the nearest preceding sunspot minimum, and their corresponding standard errors.
Also shown in the figure are the dependence of the latitudinally mean values of rotation rates  on the phase of the solar cycle relative to sunspot maximum, and their corresponding standard errors. As the figure displays,
the special feature of $B$ can also be clearly seen for the latitudinally mean values of rotation rates.

\subsection{PERIODICITY OF SOLAR DIFFERENTIAL ROTATION}
We calculate the autocorrelation coefficient
of the coefficient $A$, varying with  relative phase shifts
of $A$ with respect to itself, which is shown in Figure 5. The autocorrelation coefficient peaks to be 0.5812 when phase shift is 74 CRs or 66 months (1 CR=0.894 month, please see Li et al. (2007)), which is statistically significant at the $99.9\%$ confidential level. Thus, there is a period of about 66 months for $A$.
Similarly, we calculate the  autocorrelation coefficient
of the coefficient $B$, varying with  relative phase shifts, which is also shown in Figure 5. The autocorrelation coefficient of $B$ peaks to be 0.7931 when phase shift is 141
CRs or 126 months, which is statistically significant at the $99.9\%$ confidential level. Therefore, there is a period of about 126 months for $B$.

We calculate the  autocorrelation coefficient
of monthly mean sunspot numbers at the time interval of August 1976 to April 2008, varying with  phase shifts. Monthly mean sunspot numbers are downloaded from from the SIDC's web site\footnote{http://sidc.oma.be/sunspot-data/}, and they are found to have a period of 121 months, slightly different from that of $B$.

The monthly value of the coefficient $B$ is
obtained with linear interpolation to the two values of $B$ at its nearest two Carrington rotations. For example, the value ($B_{t}$) of $B$  at a certain month ($t$) is obtained through linear interpolation of the two values ($B_{CR1}$ and $B_{CR2}$) of $B$ at its neighbor two Carrington rotations ($CR1$ and $CR2$):
$$
B_{t}={{B_{CR1}-B_{CR2}}\over{CR1-CR2}}\times (t-CR1)+B_{CR1}.
$$
The obtained monthly values of $B$  are shown in Figure 6 together with
the original data. As the figure indicates, the interpolated
values fit the original data quite well.
Figure 7 shows the result of the cross-correlation analysis of the monthly values of $B$ and the monthly sunspot numbers, in which the
abscissa is the shift of the former
versus the latter with backward
shifts given minus values. The figure indicates that the cross-correlation coefficient has three peaks and three valleys within the shifts of $\pm 200$ months, and the time interval between neighboring two peak points
is 126 and 120 months, and that between neighboring two valley points is 123 and 120 months. The reason why the time intervals are different from one another is that the coefficient $B$ and sunspot numbers have different periods.
The cross-correlation coefficient doesn't have a maximum or minimum value when the two have no shift, indicating
that the two have a phase difference. In sum, the periodicity of the two is slightly different in period length, but obviously different in period phase.

\section{CONCLUSIONS AND DISCUSSIONS}
The solar differential rotation has a cyclic pattern of
change. This pattern can be described as a
torsional oscillation, in which the solar rotation is periodically
speeded up or slowed down in certain zones of latitude
while elsewhere the rotation remains essentially steady (Snodgrass $\&$ Howard 1985; Li et al. 2008). The
zones of anomalous rotation move on the Sun in wavelike
fashion, keeping pace with and flanking the zones of
magnetic activity (LaBonte $\&$ Howard, 1982; Snodgrass $\&$ Howard, 1985).
The surface torsional pattern, and perhaps the magnetic activity
as well, are only the shadows of another unknown phenomenon
occurring within the convection zone (Snodgrass 1987; Li et al. 2008).
In this study, the solar-cycle related variation of solar differential rotation
can be explained  on the framework of the surface torsional pattern and  the magnetic activity as follows.

Brajsa,  Ruzdjak $\&$ Wohl(2006) investigated  solar-cycle related variations of solar rotation, and they found  a higher than average rotation velocity in the minimum of activity. Then they gave a possible interpretation:
when magnetic fields are weaker, one can expect a more pronounced
differential rotation yielding a higher rotation velocity at low latitudes on an average.
Indeed as Figure 4 shows, more pronounced differentiation of rotation rates appears at the ascending part of a Schwabe cycle than at the descending part on an average, due to weaker magnetic fields appearing at the ascending part.
Strong magnetic fields repress differentiation of solar rotation, so that  rotation rates show less pronounced differentiation. But weak magnetic fields seem to just reflect differentiation of rotation rates, and further, weak magnetic fields may more effectively reflect differentiation at low latitudes with high rotation rates than at high latitudes with low rotation rates. As Figure 4 shows, $B$ may be divided into three parts. Part one spans from the start to the $4^{th}$ year of a Schwabe cycle, the absolute $B$ is approximately a constant or slightly fluctuates (the absolute $B$ seems to slightly increase for that weak magnetic fields more clearly reflect differentiation at low latitudes than at high latitudes). Part two spans from
the $4^{th}$  to the $7^{th}$ year, the absolute $B$ decreases. When solar activity is progressing into this part, sunspots appear at lower and lower latitudes, magnetic fields repress differentiation more and more effectively, and differentiation appears less and less conspicuously, thus the absolute $B$ decreases within this part. Part three spans from the $7^{th}$ year to the end of a Schwabe cycle.  Within this part, magnetic fields becoming more and more weak, repress differentiation  less and less effectively, and sunspots  appearing at more and more low latitudes lead to that the differentiation reflected by latitudinal migration should be more and more conspicuous, thus, the absolute $B$ increases.

When solar activity is progressing into a new Schwabe cycle from its former old cycle, more and more sunspots of the new Schwable cycle appear at high latitudes with relative low rotation rates, and less and less sunspots of the former old Schwable cycle
appear at low latitudes with relative high rotation rates, rotation rate of magnetic fields thus decreases when solar activity transfers to a new Schwabe cycle from its former old cycle (see figure 4).
From the $1^{st}$ to the $3.5^{th}$ of a new Schwabe cycle, rotation rate slightly increases (see Figure 4), and magnetic fields can more and more clearly reflect differentiation, namely, differentiation increases (see Figure 4),  due to the migration of sunspot towards low latitudes.
Rotation rates increase from the $3.5^{st}$ to the $7^{th}$ year, but decrease  from the $7^{th}$ year to the end of a Schwabe cycle, and the reason why rotation rates vary in such a way is the same as for $B$ given above.

In sum, the solar-cycle related variation of solar differential rotation
is inferred to be the result of both the latitudinal migration of the surface torsional pattern and the repression of strong magnetic activity. This means that measurements of differential rotation should different at different phase of a Schwabe cycle or/and at different latitudes (different spacial positions of observed objects on the solar disk). That is the main reason why many different results about solar differential rotation exist  at the present
(Howard 1984; Schroter 1985; Snodgrass 1992; Beck 1999;
Paterno 2010). The most accepted rotation law deduced through
the tracer method was derived by Newton $\&$ Nunn (1951) and based on almost 70 year of
sunspot observations: $\omega (\phi)=14.38-2.44 sin^{2}\phi \ (deg\ day^{-1})$ (Paterno 2010). The most accepted rotation law deduced from spectroscopic observations was derived by Howard $\&$ Harvey (1970):
$\omega (\phi)=13.76-1.74 sin^{2}\phi-2.19 sin^{4}\phi  \ (deg\ day^{-1})$ (Paterno 2010). If the
standard formula of differential rotation is also utilized to fit the rotation law deduced
from spectroscopic observations, the rotation law should then become:
$\omega (\phi)=14.04-3.94 sin^{2}\phi \ (deg\ day^{-1})$. Due to the repression of sunspots' magnetic fields
 to differentiation, a less pronounced
differential rotation  should appear in the tracer method, namely, the absolute value of B should be smaller for
the tracer method than for the spectroscopic method. This is inferred to be the reason why the
rotation laws deduced by the two kinds of measurement methods are different from each other.
That is to say, the rotation law deduced from spectroscopic measurements should reflect the real pattern of
differential rotation  in the solar atmosphere, but the rotation law deduced through
the tracer method, especially by tracers of strong magnetic filed, couldn't  reflect the real differentiation
of the solar atmosphere, due to the repression of sunspots' magnetic fields
 to differentiation. This is the reason why the latitudinal shift of angular rotation rate does not appear in Figure 1, but appears in the torsional oscillation pattern (Snodgrass 1987; Li et al. 2008).

Both solar magnetic activity and latitudinal migration may be regularized as the Schwabe cycle in time domain. This is the reason why $B$ basically show the Schwabe cycle. However, latitudinal migrations of different Schwabe cycles should overlap in the space domain. For example, sunspots of a new Schwabe cycle appear at high latitudes around its minimum time, however at the same time sunspots of its former Schwabe cycle still appear on the solar disk, but at low latitudes. Sunspots of a new Schwabe cycle at high latitudes and those of its former old cycle at low latitudes
give different measurements of differential rotation around the minimum time of the new schwabe cycle, which should make $B$ to have a  slightly different period in length from the Schwabe cycle. The absolute $B$  usually peaks several years after the maximum time of a sunspot activity cycle, that is  the reason why the periodicity of $B$ is  different in period phase from that of sunspot activity.

\section*{Acknowledgments}
We thank the anonymous referees for their careful reading of the
manuscript and constructive comments which improved the original
version of the manuscript.
Data of solar differential rotation rates used here are kindly given by Dr. Chu,
 to whom the authors would like to express their deep thanks.  The work is supported by the NSFC under Grants
11273057, 10921303, 11147125, and 11073010, the National Basic Research Program of China
(2011CB811406 and 2012CB957801), and the Chinese Academy of Sciences.

\newpage
\input{epsf}
\begin{figure*}
\begin{center}
\epsfysize=12.cm\epsfxsize=12.cm \hskip -5.0 cm \vskip 30.0 mm
\epsffile{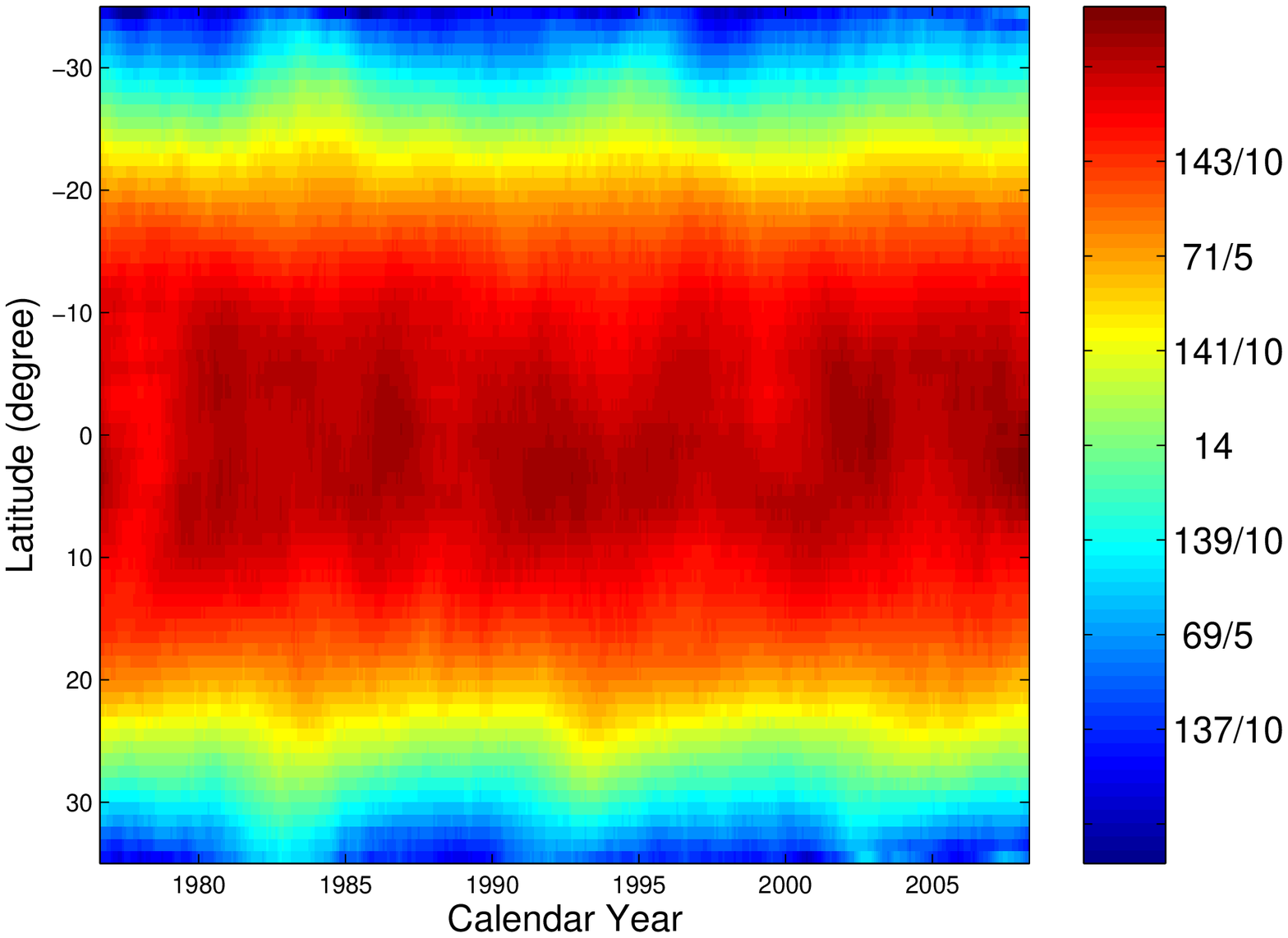} \vskip 2.5cm {{\bf Figure.1}\
The rotation angular velocities of magnetic fields distributed along latitudes and varying with time, which are obtained by Chu et al. (2010). Latitudes of the southern hemisphere are negative. The unit shown on the color bar is
$degrees\ day^{-1}$.
}
\end{center}
\end{figure*}

\newpage
\input{epsf}
\begin{figure*}
\begin{center}
\epsfysize=12.cm\epsfxsize=12.cm \hskip -5.0 cm \vskip 30.0 mm
\epsffile{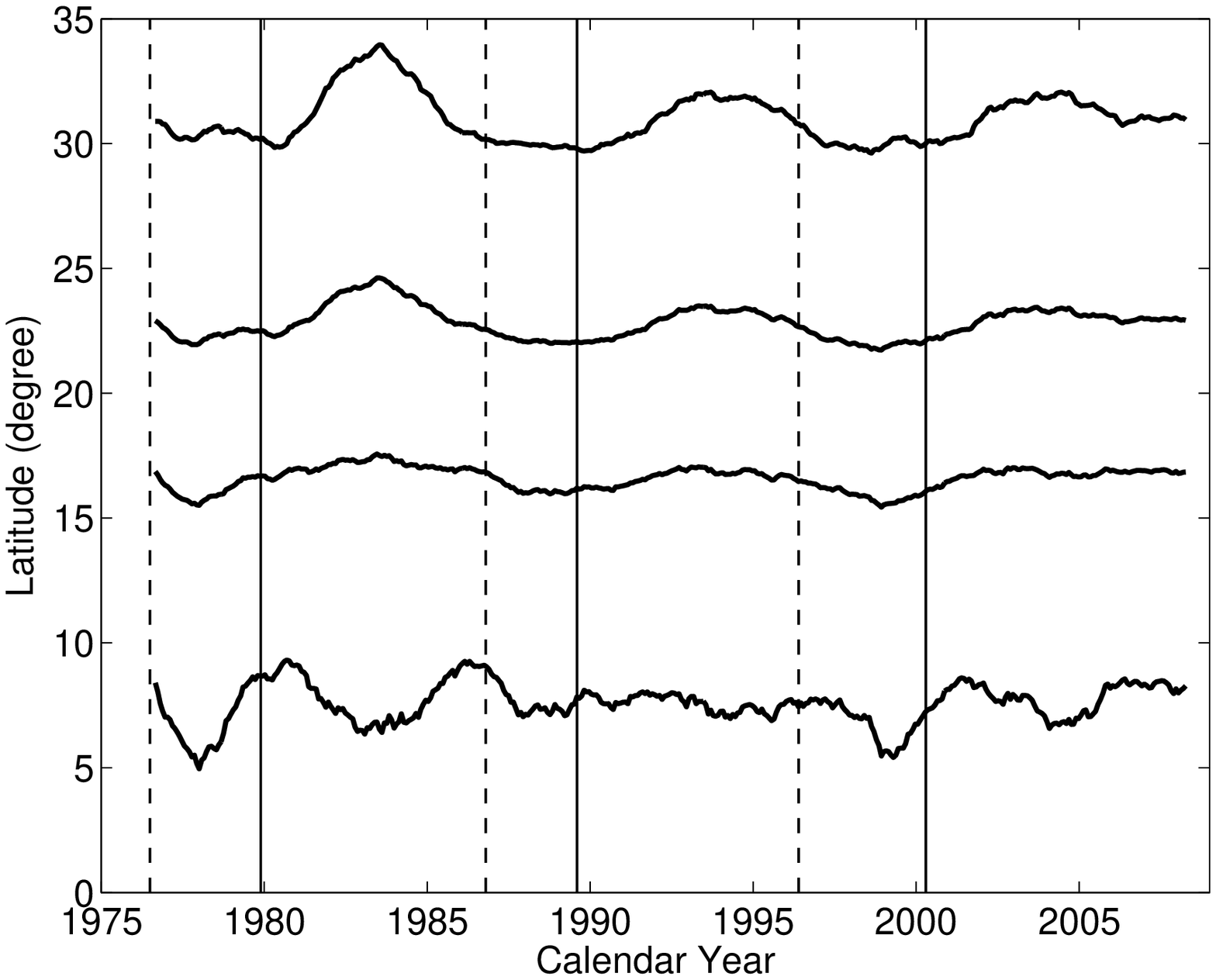} \vskip 2.5cm {{\bf Figure.2}\
Isopleth lines (the bold solid lines) of  rotation rates.  The corresponding rotation rates are 14.29, 14.25, 13.10, and $13.85^{o}\ day^{-1}$ in turn form low to high latitudes. The thin vertical dashed lines indicate the minimum times of sunspot cycles. The thin vertical solid lines indicate the maximum times of sunspot cycles.
}
\end{center}
\end{figure*}

\newpage
\input{epsf}
\begin{figure*}
\begin{center}
\epsfysize=12.cm\epsfxsize=12.cm \hskip -5.0 cm \vskip 30.0 mm
\epsffile{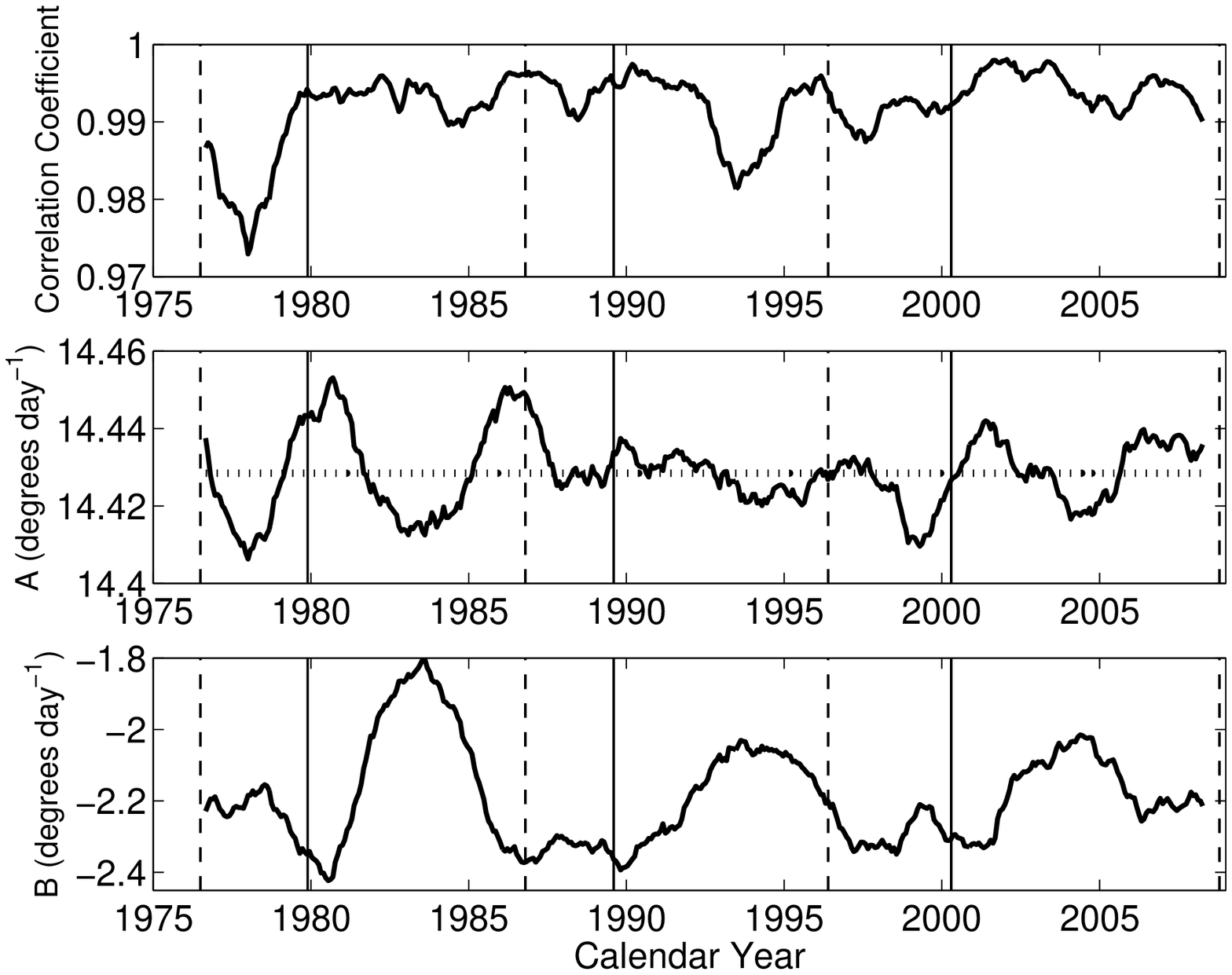} \vskip 2.5cm {{\bf Figure.3}\
The top panel: the correlation coefficient (the bold solid line)
of the latitudinal distribution of rotation rates
at each CR of the considered time interval fitted by
the standard formula of solar differential rotation.
The middle panel: the coefficient $A$ (the bold solid line) of solar differential rotation.
The horizontal dotted lines indicates the mean ($14.428^{o}\ day^{-1}$) of $A$.
The bottom panel: the coefficient $B$ (the bold solid line) of solar differential rotation.
The thin vertical dashed lines indicate the minimum times of sunspot cycles. The thin vertical solid lines indicate the maximum times of sunspot cycles.
}
\end{center}
\end{figure*}

\newpage
\input{epsf}
\begin{figure*}
\begin{center}
\epsfysize=12.cm\epsfxsize=12.cm \hskip -5.0 cm \vskip 30.0 mm
\epsffile{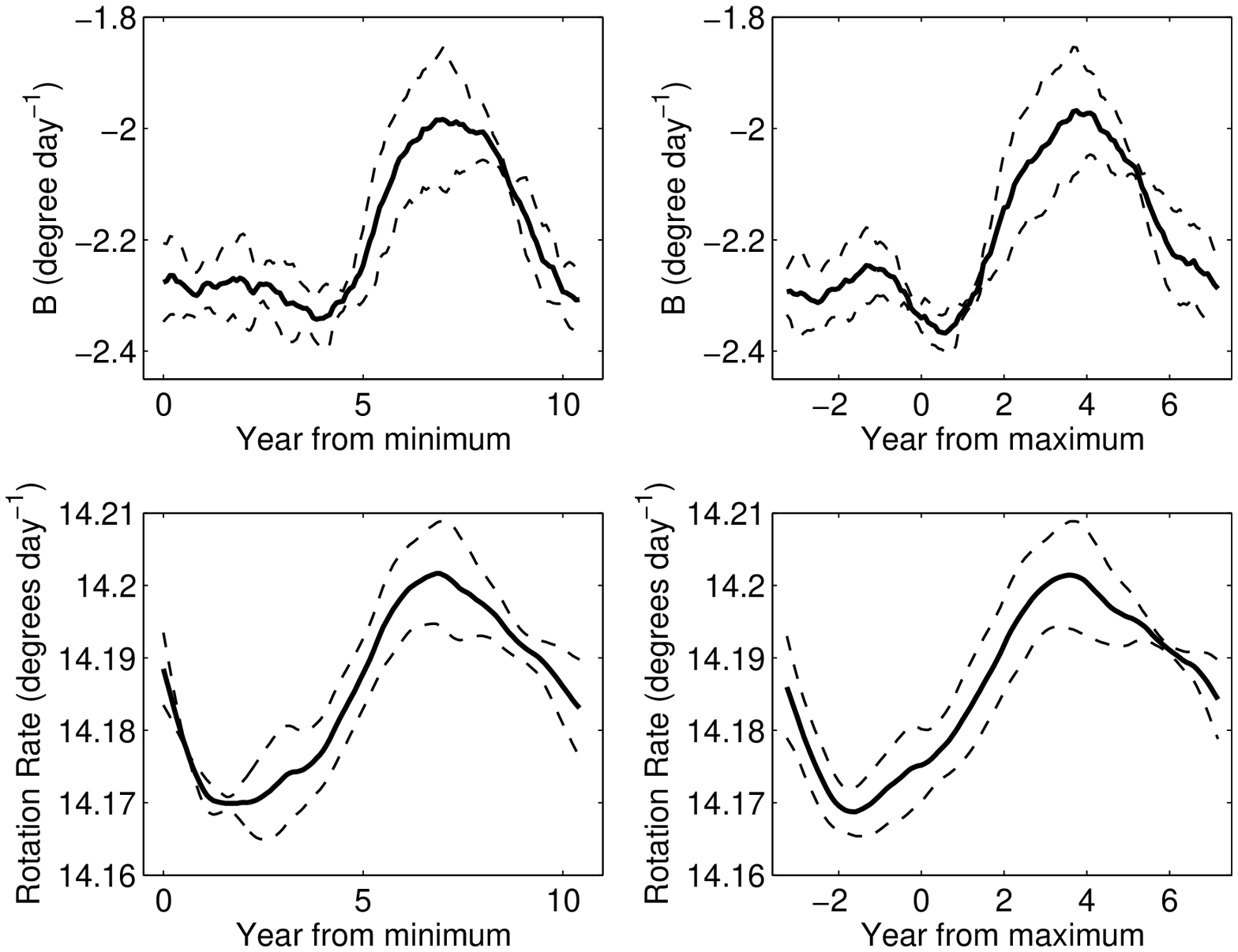} \vskip 2.5cm {{\bf Figure.4}\
Top left panel: the dependence (the solid line) of  values of the coefficient $B$ on the phase of the solar
cycle relative to the nearest preceding sunspot minimum, and their corresponding standard errors (the dashed lines).
Top right panel: the dependence (the solid line) of  values of the coefficient $B$ on the phase of the solar
cycle relative to sunspot maximum, and their corresponding standard errors (the dashed lines).
Bottom left panel: the dependence (the solid line) of  the latitudinally mean values of rotation rates  on the phase of the solar cycle relative to the nearest preceding sunspot minimum, and their corresponding standard errors (the dashed lines).
Bottom right panel: the dependence (the solid line) of  the latitudinally mean values of rotation rates  on the phase of the solar cycle relative to sunspot maximum, and their corresponding standard errors (the dashed lines).
}
\end{center}
\end{figure*}

\newpage
\input{epsf}
\begin{figure*}
\begin{center}
\epsfysize=12.cm\epsfxsize=12.cm \hskip -5.0 cm \vskip 30.0 mm
\epsffile{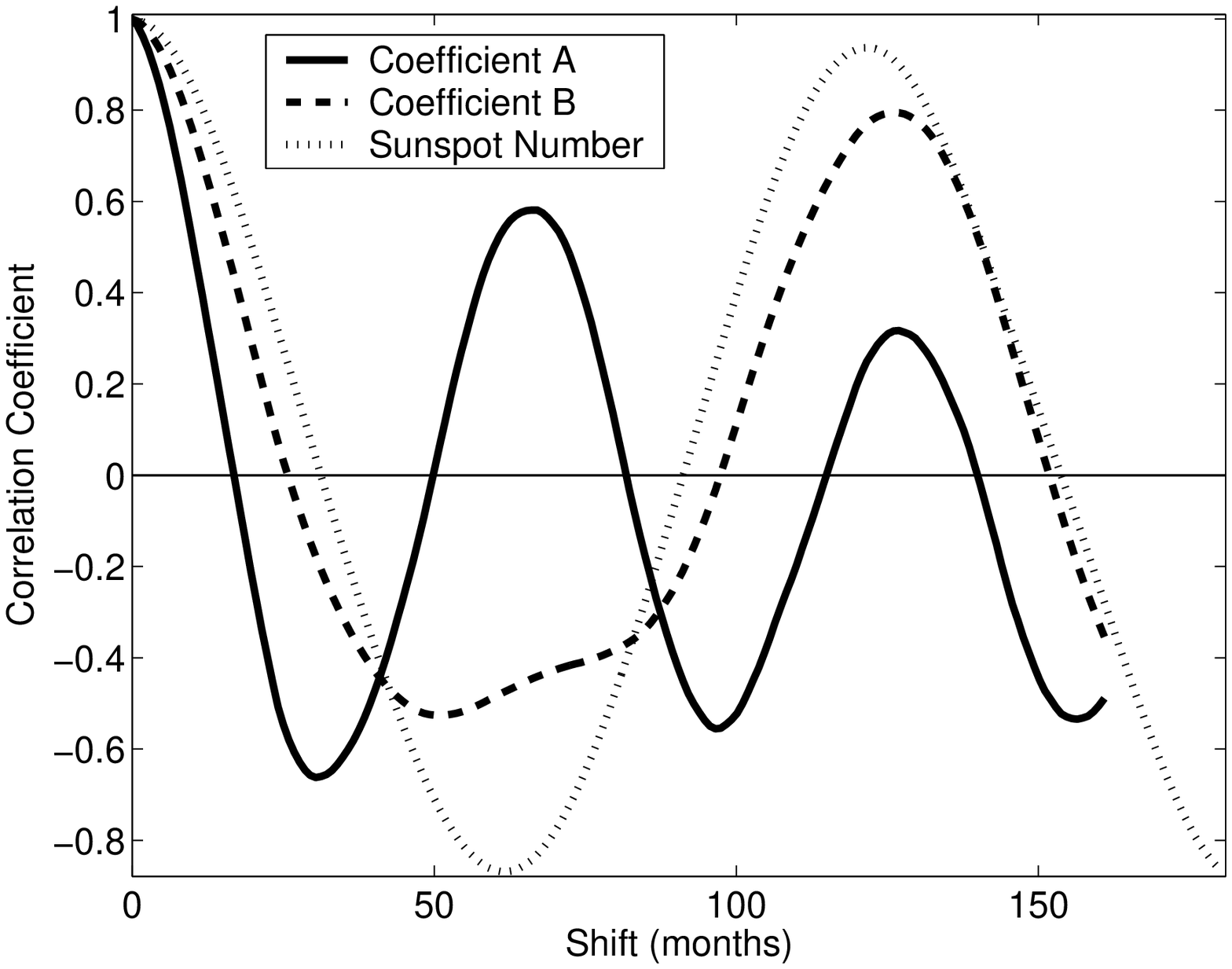} \vskip 2.5cm {{\bf Figure.5}\
Autocorrelation coefficient
of the coefficients $A$ (the bold solid line) and $B$ (the dashed line ) of rotation rates, varying with  relative phase shifts
of themselves, respectively. The dotted line shows autocorrelation coefficient
of monthly mean sunspot, with  relative phase shifts
of itself.
}
\end{center}
\end{figure*}

\newpage
\input{epsf}
\begin{figure*}
\begin{center}
\epsfysize=12.cm\epsfxsize=12.cm \hskip -5.0 cm \vskip 30.0 mm
\epsffile{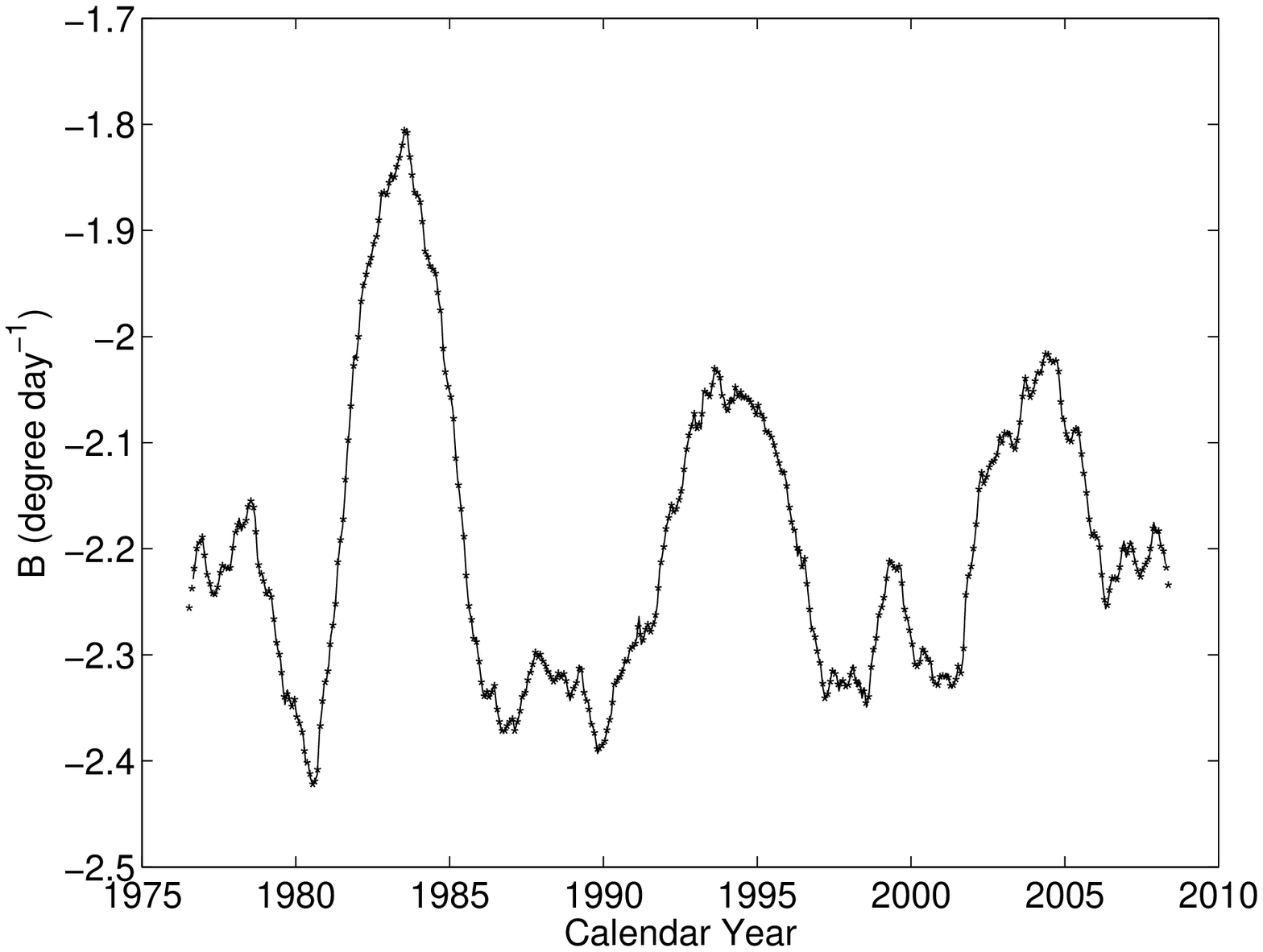} \vskip 2.5cm {{\bf Figure.6}\
Comparison of the  monthly values (asterisks) of the coefficient $B$ of differential rotations  with
the original data (the solid line).
}
\end{center}
\end{figure*}

\newpage
\input{epsf}
\begin{figure*}
\begin{center}
\epsfysize=12.cm\epsfxsize=12.cm \hskip -5.0 cm \vskip 30.0 mm
\epsffile{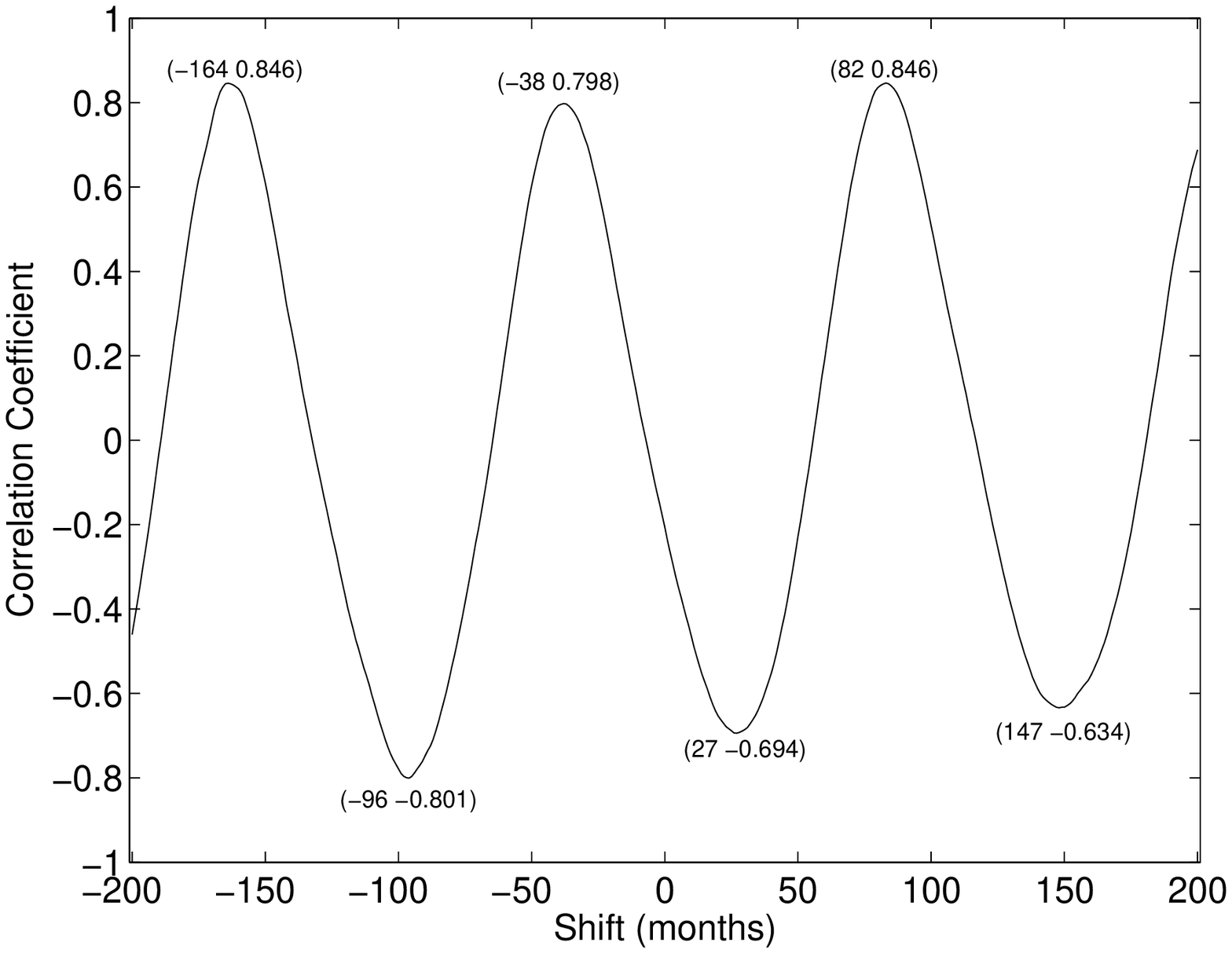} \vskip 2.5cm {{\bf Figure.7}\
The cross-correlation coefficient of the monthly values of $B$ and the monthly sunspot numbers, varying with shifts of the former
versus the latter with backward
shifts given minus values. The coordinate values of three peaks and three valleys are given in the figure.
}
\end{center}
\end{figure*}

\end{document}